\documentclass[%
reprint,
amsmath,amssymb,
aps,
superscriptaddress
]{revtex4-1}

\usepackage[utf8]{inputenc}
\usepackage{siunitx}
\usepackage{xcolor}
\newcommand{\eqnref}[1]{(\ref{#1})}
\newcommand{\us}{\si{\micro\second}}

\newcommand{\ZII}{{Z\!I\!I}}
\newcommand{\ZXI}{{Z\!X\!I}}

\newcommand{\IIZ}{{I\!I\!Z}}
\newcommand{\IXZ}{{I\!X\!Z}}

\usepackage{comment}
\usepackage{graphicx}% Include figure files
\usepackage{dcolumn}% Align table columns on decimal point
\usepackage{bm}% bold math
\usepackage{physics}
\usepackage{url}
\begin{document}

%\preprint{APS/123-QED}

\title{Three-qubit Parity Gate via Simultaneous Cross Resonance Drives}% Force line breaks with \\
%\thanks{A footnote to the article title}

\author{Toshinari Itoko}
\email{itoko@jp.ibm.com}
\affiliation{IBM Quantum, IBM Research – Tokyo, Tokyo, Japan}
\author{Moein Malekakhlagh}%
\affiliation{IBM Quantum, IBM Thomas J. Watson Research Center, Yorktown Heights, NY 10598}
\author{Naoki Kanazawa}%
\affiliation{IBM Quantum, IBM Research – Tokyo, Tokyo, Japan}
\author{Maika Takita}%
\affiliation{IBM Quantum, IBM Thomas J. Watson Research Center, Yorktown Heights, NY 10598}

% \collaboration{CLEO Collaboration}%\noaffiliation

\date{\today}% It is always \today, today,

\begin{abstract}
		Native multi-qubit parity gates have various potential quantum computing applications, such as entanglement creation, logical state encoding and parity measurement in quantum error correction. Here, using simultaneous cross-resonance drives on two control qubits with a common target, we demonstrate an efficient implementation of a three-qubit parity gate. We have developed a calibration procedure based on the one for the echoed cross-resonance gate. We confirm that our use of simultaneous drives leads to higher interleaved randomized benchmarking fidelities than a naive implementation with two consecutive CNOT gates. We also demonstrate that our simultaneous parity gates can significantly improve the parity measurement error probability for the heavy-hexagon code on an IBM Quantum processor using seven superconducting qubits with all-microwave control.

	% \begin{description}
		% Secondary publications and information retrieval purposes.
		% \item[PACS numbers]
		% May be entered using the \verb+\pacs{#1}+ command.
		% You may use the \texttt{description} environment to structure your abstract;
		% use the optional argument of the \verb+\item+ command to give the category of each item. 
		% \end{description}
\end{abstract}

% \pacs{Valid PACS appear here}% PACS, the Physics and Astronomy
\maketitle

%\tableofcontents

%%%%%%%%%%%%%%%%%%%%%%%% Sec: Revised Introduction %%%%%%%%%%%%%%%%%%%%%%%%%%%%%
\section{Introduction}
Standard implementation of quantum computing \cite{nielsen2001quantum, kitaev2002classical} involves expressing multi-qubit operations in terms of a universal set of single- and two-qubit gates \cite{barenco1995elementary}. Through quantum circuit optimization, one can achieve an equivalent shallower-depth circuit, benefiting not only from less incoherent error, caused by energy relaxation and dephasing \cite{gardiner2004quantum,clerk2010introduction, krantz2019quantum}, but also possibly from less coherent (control) error. At a high level, strategies for circuit optimization can be software- \cite{maslov2005quantum, maslov2008quantum, kliuchnikov2013asymptotically, amy2014polynomial,nam2018automated} and/or hardware-inspired \cite{burkard1999physical, sorensen2000entanglement, fedorov2012implementation, martinez2016compiling, feng2020quantum, lu2022multipartite, gu2021fast, warren2023extensive, kim2022high}: the former employs unitary group identities for simplification, while the latter considers the hardware connectivity, and explores the hardware potential to achieve more efficient two- or multi-qubit gates. Here, following the latter approach, and inspired by Cross-Resonance (CR) \cite{paraoanu2006microwave, rigetti2010fully, Tripathi_Operation_2019, Magesan_Effective_2020, Malekakhlagh_First-Principles_2020} quantum processors provided by IBM, we study a Three-qubit Parity (TP) gate~\footnote{In this term, three-qubit refers gate (not parity), in fact, TP gate checks two-qubit parity.}, and provide an efficient calibration based on the existing Echoed Cross-Resonance (ECR) scheme \cite{corcoles2013process, Sheldon2016, jurcevic2021demonstration, Malekakhlagh_First-Principles_2020, sundaresan2020reducing}.
\begin{figure}[h]
\hspace{-5mm}
	\includegraphics[width=0.9\linewidth]{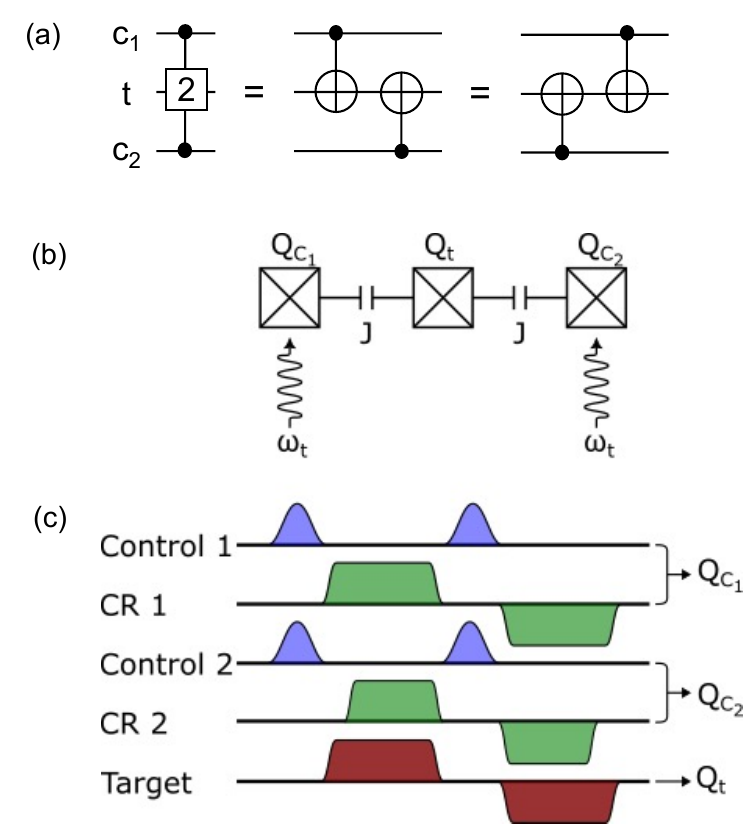}
	\caption{
		\textbf{Implementation of three-qubit parity gate with simultaneous CR drives.} \textbf{(a)} Circuit representation of a $Z$-parity gate, equivalent to two consecutive CNOTs with a common target. The case with a common control is locally equivalent up to single-qubit Hadamard gates. \textbf{(b)} Cross-resonance schematics with two control ($c_1$ and $c_2$) and one target ($t$) qubits.
		\textbf{(c)} Pulse-level implementation using simultaneous CR drives following the ECR calibration \cite{corcoles2013process, Sheldon2016, jurcevic2021demonstration, Malekakhlagh_First-Principles_2020, sundaresan2020reducing}. The CR (green) and rotary (red) pulses have a carrier frequency resonant with the target qubit.
	}
	\label{fig:idea}
\end{figure}

Having efficient parity gates \cite{reagor2022hardware, dodge2023hardware} in the native gate set is useful for numerous applications. In particular, the utility of a TP gate boils down to its local equivalence with two consecutive CNOTs (Two-CX) on three qubits, in which they share either a common control (or target) qubit [Fig.~1(a)]. Such a circuit subroutine appears for instance in (i) the creation of multi-qubit entanglement, in particular the Greenberger-Horne-Zeilinger (GHZ) state \cite{greenberger1989going, bouwmeester1999observation}, (ii) logical encoder and parity check syndrome measurement in Quantum Error Correction (QEC) \cite{shor1995scheme, calderbank1996good, terhal2015quantum}, and (iii) successive swaps across a qubit network \cite{lloyd1993potentially, Divincenzo2000universal}.

In this paper, we present a TP gate implementation that fits well with IBM's CR architecture. Our implementation closely follows that of the ECR gate \cite{Sheldon2016, jurcevic2021demonstration, Malekakhlagh_First-Principles_2020, Sundaresan2022}, but instead employs two simultaneous CR drives with a common target qubit, hence named Simultaneous Cross Resonance Parity (SCRP) gate [Fig.~\ref{fig:idea}(b)--(c)]. This protocol implements a three-qubit $Z$-parity gate, which is locally equivalent to any other TP gates. Our use of simultaneous drives should work in principle if each CR pulse leads only to a $ZX$ interaction between the intended qubits.
In other words, SCRP gives $ZXI$ and $IXZ$ interactions
\footnote{The ordering of qubits is (control-0, target, control-1) throughout this paper},
that are commutative, hence additive.
Intuitively, the SCRP implementation should improve the fidelity of the $Z$-parity gate
mainly due to its shorter pulse schedule.
We confirm that unwanted cross-drive contributions are indeed higher-order effects, and hence weaker, by deriving an effective three-qubit gate Hamiltonian using Schrieffer-Wolff Perturbation Theory (SWPT) \cite{schrieffer1966relation, Magesan_Effective_2020, Malekakhlagh_First-Principles_2020, Malekakhlagh_Mitigating_2022, malekakhlagh2022time, heya2023floquet} (Sec.~\ref{sec:HamAnalysis}). Using Interleaved Randomized Benchmarking (IRB) \cite{emerson2005scalable, knill2008randomized, magesan2011scalable, magesan2012efficient}, we demonstrate improved Error Per Gate (EPG) for the SCRP implementation compared to Two-CX. Furthermore, we demonstrate the SCRP implementation improves the fidelity of parity measurement on an IBM Quantum processor~\cite{IBMQuantum}, namely \texttt{ibm\_auckland}.
In particular, the SCRP implementation can reduce the average syndrome error probability of $X$-parity measurement for the heavy-hexagon code \cite{chamberland2020topological, chen2022calibrated, sundaresan2023demonstrating} by up to 28 percents comparing with a naive implementation with CNOT gates on the device (Sec.~\ref{sec:experiments}).

The rest of this paper is organized as follows: first, in Sec.~\ref{sec:HamAnalysis}, we study effective gate interactions for the SCRP gate implementation using SWPT. In Sec.~\ref{sec:calibration}, we discuss the SCRP calibration of the TP gate, and provide IRB results that demonstrate improvement in EPG with respect to the standard Two-CX implementation. Furthermore, in Sec.~\ref{sec:experiments}, We showcase the SCRP gate's utility in improving the syndrome measurement success probability of the heavy-hexagon code. Finally, Sec.~\ref{sec:conclusion}
concludes the paper, and examines further potential applications and extensions of the SCRP idea.

\section{Hamiltonian Analysis}
\label{sec:HamAnalysis}

We next provide a Hamiltonian analysis for the SCRP gate, based on SWPT \cite{Magesan_Effective_2020, Malekakhlagh_First-Principles_2020, Malekakhlagh_Mitigating_2022}. Our analysis clarifies why the SCRP gate works in practice: at sufficiently weak CR drive, the effective $ZXI$ and $IXZ$ rates depend only on their corresponding drive amplitudes. Furthermore, undesired three-qubit cross interactions such as the $ZXZ$ term appears only at higher order, and hence are weaker.

We model the transmon qubits as a set of Duffing oscillators with nearest-neighbor exchange interaction under Rotating-Wave Approximation (RWA) as:
\begin{align}
	\begin{split}
		\hat{H}_s & = \sum\limits_{j=c_1,c_2,t} \left(\omega_j \hat{a}_j^{\dag}\hat{a}_j + \frac{\alpha_j}{2}\hat{a}_j^{\dag}\hat{a}_j^{\dag}\hat{a}_j\hat{a}_j\right) \\
		& + \sum\limits_{\langle j,k\rangle} J_{jk} \left(\hat{a}_j^{\dag}\hat{a}_k+\hat{a}_j\hat{a}_k^{\dag}\right) \;,
	\end{split}
	\label{eq:EffHam_Hs}
\end{align}
with $\omega_j$, $\alpha_j$ and $J_{jk}$ as the qubit frequency, anharmonicity, and pairwise exchange interaction, respectively, for $j, k\in \{c_1,c_2,t\}$. Furthermore, we model the CR and a possible direct target drives as
\begin{align}
	\begin{split}
		\hat{H}_d(t) = \sum\limits_{j=c_1,c_2,t} & \frac{1}{2}\Big[ \Omega_{j}^{*}(t)e^{i\omega_{d} t} \hat{a}_j + \Omega_{j}(t)e^{-i\omega_{d} t } \hat{a}_j^{\dag} \Big] \;,
	\end{split}
	\label{eq:EffHam_Hd}
\end{align}
with $\Omega_{j}(t)\equiv \Omega_{jX}(t)+i\Omega_{jY}(t)$ and $\omega_{d}$ denoting the complex-valued envelope and the common carrier frequency, respectively. In the rotating frame (RF) of the drive, which is set to the target qubit frequency, the Hamiltonian simplifies to:
\begin{align}
	\begin{split}
		\hat{H}_{\text{rf}} (t) & \equiv \sum\limits_{j=c_1,c_2,t} \left(\Delta_{jd} \hat{a}_j^{\dag}\hat{a}_j + \frac{\alpha_j}{2}\hat{a}_j^{\dag}\hat{a}_j^{\dag}\hat{a}_j\hat{a}_j\right) \\
		& + \sum\limits_{\langle j,k\rangle} J_{jk} \left(\hat{a}_j^{\dag}\hat{a}_k+\hat{a}_j\hat{a}_k^{\dag}\right) \\
		& + \sum\limits_{j=c_1,c_2,t} \frac{1}{2}\Big[\Omega_{j}^{*}(t)\hat{a}_j+\Omega_{j}(t)\hat{a}_j^{\dag}\Big] \;.
	\end{split}
	\label{eq:EffHam_Hrf}
\end{align}
where $\Delta_{jd}\equiv \omega_{j}-\omega_{d}$. The RF Hamiltonian~(\ref{eq:EffHam_Hrf}) is the starting point of our analysis. To understand the SCRP power budget, for simplicity, we assume an always-on X-quadrature-only CW drive $\Omega_{j}(t)=\Omega_{j}$.

Applying time-independent SWPT, we derive effective (resonant) interactions for the SCRP gate through recursive frame transformations that averages over off-resonant transitions \cite{Magesan_Effective_2020, Malekakhlagh_First-Principles_2020, Malekakhlagh_Mitigating_2022}. The relevant SCRP frame is diagonal with respect to the two control qubits, i.e. allowing only $I$ and $Z$ on the controls, and off-diagonal with respect to the target. We treat the first two lines of Eq.~(\ref{eq:EffHam_Hrf}) as the bare, and the last line as the interaction Hamiltonian.

Up to the zeroth order, the exchange interaction leads to nearest-neighbor static $ZZ$ interactions:
\begin{align}
	& \omega_{ZZI}^{(0)} = \frac{J_{c_1t}^2} {\Delta_{c_1t}-\alpha_t} - \frac{J_{c_1t}^2} {\Delta_{c_1t}+\alpha_{c_1}} \;,
	\label{eq:EffHam_ZZI_0th}\\
	& \omega_{IZZ}^{(0)} = \frac{J_{c_2t}^2} {\Delta_{c_2t}-\alpha_{c_2}} - \frac{J_{c_2t}^2} {\Delta_{c_2t}+\alpha_{c_2}}  \;.
	\label{eq:EffHam_IZZ_0th}
\end{align}
Up to the dominant (linear) order in drive amplitudes, the $ZXI$ and $IXZ$ terms are independent, i.e. no cross-drive exists, justifying why such a simultaneous calibration works:
\begin{align}
	& \omega_{ZXI}^{(1)} = -\frac{J_{c_1t} \alpha_{c_1}}{\Delta_{c_1t}(\Delta_{c_1t}+\alpha_{c_1})} \Omega_{c_1} \;,
	\label{eq:EffHam_ZXI_1st}\\
	& \omega_{IXZ}^{(1)} = -\frac{J_{c_2t} \alpha_{c_2}}{\Delta_{c_2t}(\Delta_{c_2t}+\alpha_{c_2})} \Omega_{c_2} \;,
	\label{eq:EffHam_IXZ_1st}\\
	& \omega_{IXI}^{(1)} = \Omega_{t} -\frac{J_{c_1t} }{\Delta_{c_1t}+\alpha_{c_1}}\Omega_{c_1} -\frac{J_{c_2t} }{\Delta_{c_2t}+\alpha_{c_2}}\Omega_{c_2} \;.
	\label{eq:EffHam_IXI_1st}
\end{align}
At second, and higher-order in drive amplitudes, we find cross-drive contributions to the Stark shifts, $ZZI$ and $IZZ$ rates, as well as to $ZXI$, $IXZ$ and $ZXZ$ terms.

Based on Eqs.~(\ref{eq:EffHam_ZXI_1st})-(\ref{eq:EffHam_IXZ_1st}), the cross-drive-free nature of the desired $ZXI$ and $IXZ$ rates up to the leading order allows us to employ the existing CR echo calibration \cite{corcoles2013process,Sheldon2016, Malekakhlagh_First-Principles_2020,jurcevic2021demonstration, sundaresan2020reducing} in constructing the SCRP gate. In particular, the CR echo sequence removes the $IXI$ term, and suppresses the dominant error terms $ZZI$ and $IZZ$ up to the leading order. We discuss the SCRP calibration in more detail in the following section.

\section{Parity gate calibration}
\label{sec:calibration}
Our SCRP pulse schedule for implementing a three-qubit $Z$-parity gate is shown in Fig.~\ref{fig:idea}(c),
which is inspired by the CR echo calibration for a CNOT gate~\cite{corcoles2013process,Sheldon2016}.
The main part (green) consists of two echoed sequences of simultaneous CR drives onto the control qubits $c_1$ and $c_2$ with the carrier frequencies set to the target qubit frequency. Moreover, interleaving $X_\pi$ pulses (purple) onto the control qubits $c_1$ and $c_2$ allows for echoing out nearerst-neighbor $ZZ$ (i.e. $ZZI$ and $IZZ$), as well as the $IXI$ Hamiltonian terms up to the leading order~\cite{Malekakhlagh_First-Principles_2020, sundaresan2020reducing}.
Each individual CR echo calibration may also be accompanied with simultaneous resonant rotary tones onto the target qubit $t$ (shown altogether in red) \cite{sundaresan2020reducing}. The rotary tones were designed to suppress several unwanted terms in the effective Hamiltonian of the echoed CR drives, namely the $Y$ error on the target as well as target-spectator crosstalk \cite{sundaresan2020reducing}. To implement a $Z$-parity gate, three additional local Clifford instructions are needed in front (or back) of the schedule; namely $Z_{\pi/2}$ on $c_1$ and $c_2$, and $X_{\pi}$ on $t$.

We have developed a straighforward calibration procedure for the $Z$-parity gate based on the well-established CR echo calibration for CNOT gates~\cite{corcoles2013process,Sheldon2016},
where we adopt the two CR echo pulse configurations, i.e. amplitudes and angles, while we replace the independently calibrated rotary tones with a simultaneous SCRP rotary tone. For example, to implement a $Z$-parity gate on qubits (0, 1, 2), we use pulse amplitudes and angles calibrated for CR(0, 1) and CR(2, 1) as those for two echoed CR pulses to drive simultaneously. We place two echoed CR sequences so that their $X_\pi$ pulses in center are aligned as shown in Fig.~\ref{fig:idea}(c). Note that we could recalibrate those CR pulses at once so that they have the same duration and the resulting rotation in the target qubit becomes the desired angle for any binary input to the two control qubits, i.e. $\pi$ for 00, $-\pi$ for 11, and $0$ for 01 and 10.
However, for simplicity, we reuse CR pulse configurations for two-qubit gates to implement SCRP gates in all experiments we conduct hereafter.
% However, our preliminary experiments suggest the improvement by such an extra calibration should be marginal or negligible.
% Considering deployment to real systems, smaller cost of calibration is preferable.
% We thus reuse CR pulse configurations without any refinement in all experiments we conduct hereafter.
% 
To calibrate the simultaneous rotary tone in our SCRP implementation, we adopt and generalize the Hamiltonian Error Amplifying Tomography (HEAT) technique~\cite{sundaresan2020reducing} (See Appendix~\ref{app:heat}).
% Our preliminary experiments suggest the improvement by addition of a rotary tone
% is often not negligible while the extent varies depending on qubit triplets in use.
% We include a rotary tone for the SCRP implementation in all experiments described in this paper.

We characterize the potential improvement by the SCRP implementation in the fidelity of a $Z$-parity gate by comparing it with a naive implementation with two consecutive CNOT gates using IRB \cite{magesan2012efficient}. We prepare two interleaved sequences from a common reference Clifford sequence.
Both interleave a $Z$-parity gate, but with different implementations:
one implemented with SCRP and the other implemented with two CNOT gates (See Appendix~\ref{app:irb}). We conducted such an IRB experiment using qubits (8, 11, 14) on \texttt{ibm\_auckland}.
We used ten Clifford lengths: 2, 3, 4, 5, 7, 9, 12, 17, 25, 38. For each Clifford length, we sampled 50 RB circuits and computed survival rate from 400 shots for each circuit. We fit an exponential curve to the averaged survival rates (over the IRB seeds and shots) \cite{magesan2012efficient}.

\begin{figure}
	\includegraphics[width=\linewidth]{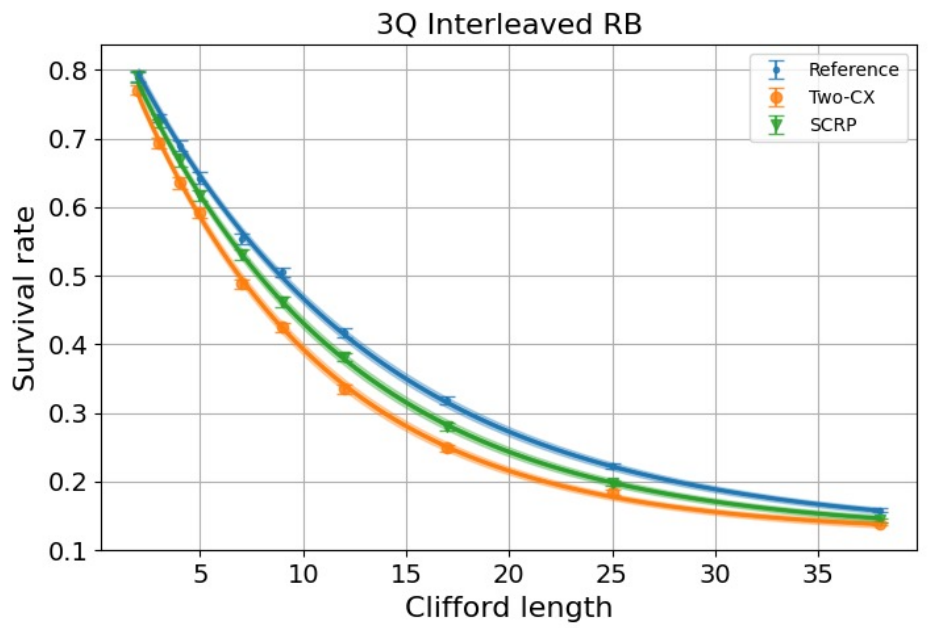}
	\caption{Interleaved RB comparing two $Z$-parity gate implementations, the consecutive two CNOT gates (Two-CX, orange) and the simultaneous CR drives (SCRP, green)
	on qubits (8, 11, 14) on \texttt{ibm\_auckland}.
	The estimated EPGs are $0.02109\pm0.00105$ (Two-CX) and $0.00964\pm0.00095$ (SCRP).}
% Error bar = standard deviation of sample mean
	\label{fig:irb}
\end{figure}

Figure~\ref{fig:irb} shows the result obtained from the IRB experiment. It contains three decay curves corresponding to
a reference sequence (blue), interleaved SCRP implementation (green), and interleaved Two-CX implementation (orange) of the $Z$-partiy gate, respectively.
% the SCRP sequence that interleaves $Z$-parity gate implemented with SCRP (green), and the Two-CX sequence that interleaves $Z$-parity gate as two consecutive CNOT gates (orange), respectively.
%Comparing two sequences interleaving DTCX and Two-CX using a common reference sequence.
The decay curve of SCRP appears clearly higher than that of Two-CX, suggesting a higher gate fidelity. For reference, the EPG estimated by the ratio of decay rates (the reference and the sequence of interest) was improved from $0.02109\pm0.00105$ (Two-CX) to $0.00964\pm0.00095$ (SCRP). This improvement is in part owed to the reduction in the gate length from 704.0 ns (Two-CX) to 369.8 ns (SCRP). Estimating best possible average gate error based on the coherence limit~\cite{gambetta2012characterization,wei2023characterizing}, we find the limits as $0.0122$ (Two-CX) and $0.00645$ (SCRP). These coherence limits are calculated from the gate lengths, $T_1$ values of $(122.7, 134.8, 159.7)$ \us, and $T_2$ values of $(73.4, 111.4, 170.3)$ \us,
for \texttt{ibm\_auckland} qubits (8, 11, 14), respectively (See Appendix~\ref{app:coherence}).
% ZZ(8, 11, 14) [dt]: DTCR=1664 <- TwoCX=3168
% ZZ(8, 11, 14) [ns]: DTCR=369.8 <- TwoCX=704.0
% CLE of ZZ(8, 11, 14): DTCR=0.006449 <- TwoCX=0.012232
% T1(8, 11, 14) [us]: ['122.7', '134.8', '159.7']
% T2(8, 11, 14) [us]: ['73.4', '111.4', '170.3']

\begin{figure*}
	\includegraphics[width=0.9\linewidth]{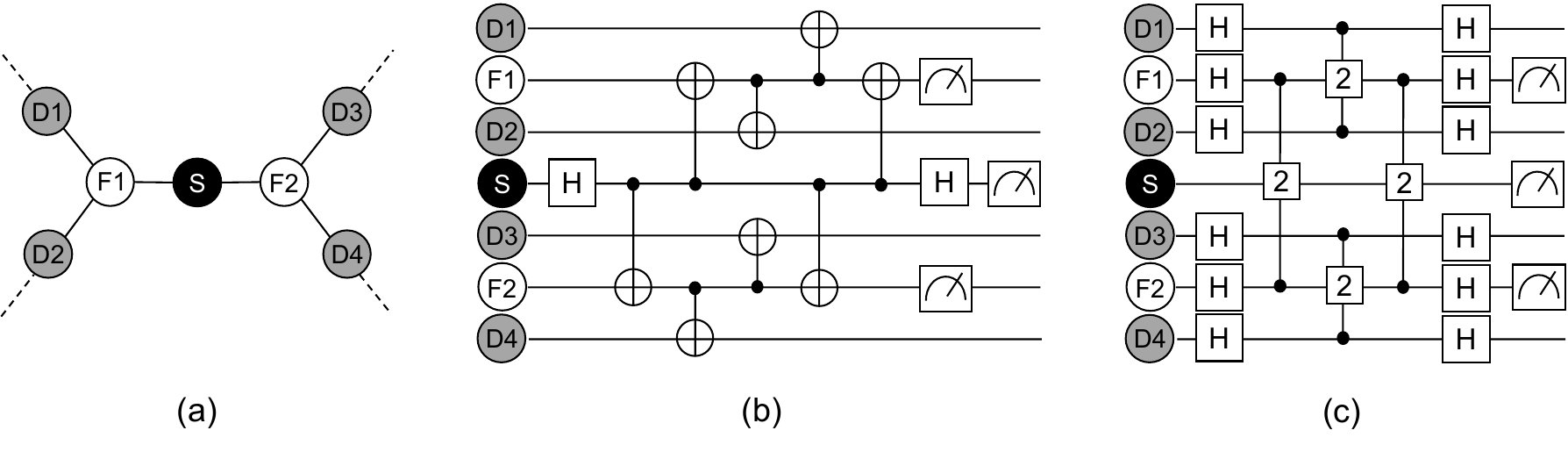}
	\caption{
		\textbf{X-Parity measurement circuit for heavy-hexagon code.}
		\textbf{(a)} 7-qubit subsystem of interest.
		\textbf{(b)} Original representation with CNOT gates~\cite{chamberland2020topological}.
		\textbf{(c)} Representation with $Z$-parity gates useful for SCRP implementation,
		which reduces the circuit depth by a factor of approximately 3/5, and uses only 4 SCRP gates compared to 8 CNOT gates.
	}
	\label{fig:heavy-hex-code}
\end{figure*}

\section{Experiments}
\label{sec:experiments}
We next demonstrate how the SCRP calibration improves the fidelity of parity measurement for QEC on IBM devices. Here, we focus on the $X$-parity measurement of the heavy-hexagon code~\cite{chamberland2020topological, Chen2021, Sundaresan2022}. The circuit realization requires seven qubits, consisting of four data qubits (D1--D4; gray), two flag qubits (F1 and F2; white) and one syndrome qubit (S; black), with a connectivity with degree at most three, as shown in Fig.~\ref{fig:heavy-hex-code}(a). The standard $X$-parity check circuit is originally represented with eight CNOT gates as shown in Fig.~\ref{fig:heavy-hex-code}(b).
It consists of four pairs of two CNOT gates with a common control and distinct target qubits, i.e. $X$-parity gates, which are locally equivalent to $Z$-parity gates up to a change of basis using single-qubit hadamard gates.
Applying the replacement, the $X$-parity check circuit will have an efficient representation with just four $Z$-parity gates as shown in Fig.~\ref{fig:heavy-hex-code}(c). We used the latter circuit representation and compared the
(i) Two-CX, and (ii) SCRP implementations of the $Z$-parity gates.

The \texttt{ibm\_auckland} processor has 27 qubits,
from which we used qubits (5, 8, 9, 11, 13, 14, 16) ordered as (D1, F1, D2, S, D3, F2, D4). 
Qubit transition frequencies ($\omega_{01}/2\pi$) of
the four data qubits (5, 9, 13, 16) are (4.99282, 5.08839, 5.01678, 4.96965) GHz,
the two flag qubits (8, 14) are (5.20360, 5.16698) GHz,
and the syndrome qubit 11 is 5.05517 GHz, respectively. The qubit anharmonicities $\alpha/2\pi$ do not vary substantially, and are approximately equal to $-340$ MHz.
% Freq/Anharmonicity (reported):
% q(5): 4.99282 GHz / -344.462 MHz
% q(8): 5.20360 GHz / -340.659 MHz
% q(9): 5.08839 GHz / -342.516 MHz
% q(11): 5.05517 GHz / -342.154 MHz
% q(13): 5.01678 GHz / -343.658 MHz
% q(14): 5.16698 GHz / -341.961 MHz
% q(16): 4.96965 GHz / -343.889 MHz

Following Sec.~\ref{sec:calibration}, we calibrated the SCRP gates on the three qubit triplets \{(5, 8, 9), (8, 11, 14), (13, 14, 16)\} found in Fig.~\ref{fig:heavy-hex-code}(c).
In advance, we also calibrated CR pulses for qubit pairs (5, 8), (9, 8), (13, 14), (16, 14),
for which the default CNOT gates are implemented with CR pulses in the opposite direction.
That means, for example, CNOT(5, 8) is implemented with CR(8, 5), i.e. CR drive on qubit 8 within the frame of qubit 5,
while CR(5, 8) is necessary to implement SCRP gate on (5, 8, 9).

We initialized the four data qubits using all possible 16 product states ranging from $\ket{+\!+\!++}$ to $\ket{-\!-\!--}$.
Here, $\ket{+}$ and $\ket{-}$ are the eigenstates of Pauli $X$ operator.
For each input state, we ran the parity check circuit 40,000 times.
We scheduled circuits in an as-late-as-possible manner, where
the total duration of the resulting circuits were 2261 ns (Two-CX) and 1365 ns (SCRP)
excluding the input state preparation and the final measurements.

We quantified how much the use of SCRP gate improves the accuracy of the parity measurement
by comparing the syndrome and the data error probabilities.
The syndrome error probability is the probability that an incorrect bit is measured at the syndrome qubit.
Here, the correct syndrome is 0 when the number of $+$ in an input state is even, and 1 when odd.
The data error probability is the probability that a state different from the input is measured at the end of a parity check circuit.
Note that those values are affected by SPAM (State Preparation And Measurement) errors. As shown in Table~\ref{tab:results},
the syndrome error probability averaged over all 16 initial states is significantly improved by the SCRP implementation from 0.1229 down to 0.0885 ($\approx$ 28\% improvement),
while the average data error rate is reduced from 0.1641 to 0.0957 ($\approx$ 42\% improvement).

\begin{table*}[tbp]
	\centering
	\caption{Syndrome and data error probabilities averaged over 16 initial states of the $X$-parity measurement.
	Individual qubit date errors are described in D1--D4 columns.}
	\label{tab:results}
	\vspace{2mm}
	\begin{tabular}{lcccccc}
		\hline
		 & Syndrome error (std) & Data error & D1 & D2 & D3 & D4 \\
		\hline
		SCRP & 0.088459 (0.001419) & 0.095697 & 0.034948 & 0.031270 & 0.017192 & 0.016798 \\
		Two-CX & 0.122878 (0.001639) & 0.164109 & 0.045063 & 0.034634 & 0.066089 & 0.030228 \\ \hline
	\end{tabular}
\end{table*}

We conducted the same experiment on different devices and qubits, and obtained similar results as above.
For example, the average syndrome error probability was improved
from 0.1566 to 0.1252 on qubits (0, 1, 2, 4, 6, 7, 10) of  \texttt{ibmq\_mumbai}
%and from 0.3014 to 0.1892 on qubits (92, 102, 101, 103, 105, 104, 111) of  \texttt{ibm\_brisbane}, respectively.
(See Appendix~\ref{app:experiments-dd} for more experimental results including the cases when running circuits with dynamical decoupling sequences).

\section{Conclusion and outlook}
\label{sec:conclusion}

We have presented a pulse-level implementation of $Z$-parity gate with simultaneous CR drives. We have shown that this SCRP implementation has little unwanted Hamiltonian terms in theory and
hence it can achieve better gate fidelity than a naive implementation with CNOT gates in practice.
We have also demonstrated using IBM CR devices that our calibrated parity gates significantly improve the error probability of the parity measurement for heavy-hexagon code.
That suggests, as the cost of SCRP gate calibration is not large,
optimizing circuits using $Z$-parity gates can be a good option for
reducing errors on superconducting quantum computing devices with all-microwave control.

Although we focused on the X-parity measurement of the heavy-hexagon code in Section~\ref{sec:experiments},
$Z$-parity gate is also naturally useful for the Z-parity measurement.
Also, our method for calibrating the $Z$-parity gate can be extended to four- or more-qubit parity gates,
which are required for other QEC code such as the surface code on a square lattice \cite{bravyi1998quantum, dennis2002topological, kitaev2006anyons, fowler2009high, fowler2012surface} or another LDPC code on more dense lattice~\cite{bravyi2023high}.

One limitation in the SCRP approach, not mentioned in Sec.~\ref{sec:experiments}, is that
CR pulses cannot be always calibrated in all pairs of coupled qubits,
e.g., due to frequency collisions in physical qubits with fixed frequencies~\cite{hertzberg2021laser,heya2023floquet}.
This suggests that for qubit triplets that are close to frequency collisions, tuning a SCRP gate might not be optimal.
We, however, expect that improvements in manufacturing process techniques such as laser annealing \cite{zhang2022high} makes our proposal more feasible. Secondly, we have assumed that cross-drive errors in $Z$-parity gate with the SCRP implementation
is negligible in our pulse strength regime based on the discussion in Sec.~\ref{sec:HamAnalysis}. This assumption, however, breaks down for faster SCRP gate implementation which requires stronger drives.

% And that may require to extend the threshold analysis of a QEC code of interest
% so that it can deal with three-qubit errors during $Z$-parity gates.
% It could be one of challenging future work.

\begin{figure}
	\includegraphics[width=0.9\linewidth]{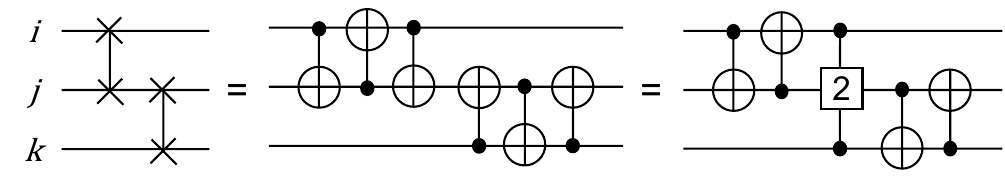}
	\caption{Optimizing the decomposition of a chain of SWAP gates using $Z$-parity gates. For $N\geq 2$ successive swaps, standard decomposition requires 3$N$ CNOTs. The $Z$-parity decomposition, however, requires $N-1$ $Z$-parity, $N+2$ CNOTs, and $O(2N)$ Hadamard gates. Assuming a similar gate time for the $Z$-parity and CNOT gates implemented via SCRP and ECR, the decomposition reduces the circuit depth by a factor of approximately $2/3$.}
	\label{fig:swap-chain}
\end{figure}

It is worth noting that
supporting a parity gate as a native instruction will be useful not only for improving parity measurements but also for optimizing circuits aimed for noisy quantum computers without QEC. For example, circuits with a chain of SWAP gates can be optimized using $Z$-parity gates.
Such circuits often appear after qubit routing, which transforms a circuit to be executable on a quantum computer with limited qubit connectivity~\cite{zulehner2018efficient,li2019tackling,itoko2020optimization,baumer2023efficient}.
As a SWAP gate is symmetric and SWAP($i, j$) is equivalent with
CNOT($i, j$) - CNOT($j, i$) - CNOT($i, j$),
two consecutive SWAP gates with a common qubit, SWAP($i, j$) and SWAP($j, k$),
can be decomposed into a sequence with a $Z$-parity gate and four CNOT gates
as shown in Fig.~\ref{fig:swap-chain}.
% CNOT($i, j$), CNOT($j, i$), $Z$-parity($i, j, k$), CNOT($j, k$), CNOT($k, j$).
The sequence using a $Z$-parity gate will have a shorter circuit length than a naive sequence with six CNOT gates, and hence should have a higher fidelity.

\section{Acknowledgment}
The authors thank Luke Govia, David Mackay, Emily Pritchett, and Xuan Wei for helpful discussions and suggestions.
For the development of the analytical techniques used to derive the effective gate-operation, MM acknowledges the support of the Army Research Office under Grant Number W911NF-21-1-0002. The views and conclusions contained in this document are those of the authors and should not be interpreted as representing the official policies, either expressed or implied, of the Army Research Office or the U.S. Government. The U.S. Government is authorized to reproduce and distribute reprints for Government purposes notwithstanding any copyright notation herein.

\appendix

\newcommand{\II}{{I\!I}}
\newcommand{\IX}{{I\!X}}
\newcommand{\IY}{{I\!Y}}
\newcommand{\IZ}{{I\!Z}}
\newcommand{\ZI}{{Z\!I}}
\newcommand{\ZX}{{Z\!X}}
\newcommand{\ZY}{{Z\!Y}}
\newcommand{\ZZ}{{Z\!Z}}
\newcommand{\IP}{{I\!P}}
\newcommand{\ZP}{{Z\!P}}
\newcommand{\QP}{{Q\!P}}
\newcommand{\IPI}{{I\!P\!I}}
\newcommand{\ZPI}{{Z\!P\!I}}
\newcommand{\IPZ}{{I\!P\!Z}}
\newcommand{\ZPZ}{{Z\!P\!Z}}
\newcommand{\QPR}{{Q\!P\!R}}

\section*{Appendix}

\subsection{Calibration of rotary tone for SCRP gate} \label{app:heat}
We describe how we calibrated the rotary tone on the target qubit for the SCRP implementation of a $Z$-parity gate.
We calibrated only the amplitude of the rotary tone in this paper, however,
our technique is applicable to calibrating the angle as well.
We swept fifty amplitude values equally spaced between $0$ and a value
which corresponds to about $X_{2\pi}$ rotation,
and set it to minimize the total estimated error.
%as shown in Fig.~\ref{fig:rotamp-cal}.
We defined a cost function for the error
$$\sum_{Q, R \in \{I, Z\}, P \in \{Y, Z\}} \|A_\QPR\|,$$
where $A_\QPR$ denotes the coefficient of a Pauli $\QPR$ in the time evolution operator for the gate (discussed later in Eq.\eqnref{eq:U_dtcr}).
In the following, we explain how to estimate the cost function from
experimentally available data following and generalizing the
HEAT (Hamiltonian Error Amplifying Tomography) technique~\cite{sundaresan2020reducing}.
Note that another generalization of HEAT to capture non-Markovian off-resonant errors, not considered in this paper, is proposed in~\cite{wei2023characterizing}.

%\begin{figure}
%	\includegraphics[width=0.88\linewidth]{fig-rotamp-cal.pdf}
%	\caption{Rotary tone calibration for echoed DTCR pulse gate.}
%	\label{fig:rotamp-cal}
%\end{figure}

\subsubsection{Echoed CR gate analysis}
We first briefly recap HEAT for echoed CR gates with rotary tones to implement $\ZX_{\pi/2}$ gate, following \cite{sundaresan2020reducing}.
HEAT was developed to characterize the time-evolution according to a block-diagonal Hamiltonian.
In the case of echoed CR gate, the time evolution unitary operator $U$ over the gate duration $t_g$ can be represented in a block-diagonal Pauli basis as:
%\begin{align} \label{eq:U_unitary}
%	U& = A_\II \II + A_\IX \IX + A_\IY \IY + A_\IZ \IZ  \nonumber \\
%	& A_\ZX \ZX + A_\ZY \ZY + A_\ZZ \ZZ.
%\end{align}
\begin{equation} \label{eq:U_unitary}
	U = \sum_{Q \in \{I, Z\}, P \in \{I, X, Y, Z\}} A_\QP\, \QP.
% 	U =  A_\II \II + \sum_{Q \in \{I, Z\}, P \in \{X, Y, Z\}} A_\QP\, \QP.
\end{equation}
%Note that there is no $\ZI$ term since it is canceled out by the echoing.
HEAT estimates the coefficients $A_{\QP}$ from experimentally available statistics.
Finally, it reconstructs the coefficients of effective Hamiltonian $\tilde{H}$ by $\tilde{H}= i \log (U) / 2t_g$.
Here, we omit the last step and use the coefficients of $U$
when using HEAT for the rotary tone calibration.
% assuming the minimization of coefficients of unwanted terms in the Hamiltonian $\tilde{H}$
% can be approximately achieved by the minimization of those in the unitary $U$.

The block-diagonal form of $U$ means that we have independent subspaces corresponding to initial control states.
If the controls is in $\ket{0}$, the evolution of the target qubit is described by
\begin{equation} \label{eq:U0}
	U_{\ket{0}} = \sum_{P \in \{I, X, Y, Z\}} A_P^{\ket{0}} P
	= \sum_{P \in \{I, X, Y, Z\}} (A_\IP + A_\ZP) P,  \nonumber
% 	= A_\II I  + \sum_{P \in \{X, Y, Z\}} (A_\IP + A_\ZP) P  \nonumber
\end{equation}
and, if the control is in $\ket{1}$, by
\begin{equation} \label{eq:U1}
	U_{\ket{1}} = \sum_{P \in \{I, X, Y, Z\}} A_P^{\ket{1}} P
	= \sum_{P \in \{I, X, Y, Z\}} (A_\IP - A_\ZP) P.  \nonumber
% 	= A_\II I  + \sum_{P \in \{X, Y, Z\}} (A_\IP - A_\ZP) P.  \nonumber
\end{equation}
%\begin{align} \label{eq:U0}
%	U_{\ket{0}} & = A_I^{\ket{0}}  + A_X^{\ket{0}}  + A_Y^{\ket{0}}  + A_Z^{\ket{0}}  \\
%	& = A_\II I + (A_\IX + A_\ZX) X + (A_\IY + A_\ZY) Y  + (A_\IZ + A_\ZZ) Z   \nonumber
%\end{align}
%while if the control is in $\ket{1}$ then
%\begin{align} \label{eq:U1}
%	U_{\ket{1}} & = A_I^{\ket{1}}  + A_X^{\ket{1}}  + A_Y^{\ket{1}}  + A_Z^{\ket{1}}  \\
%	& = A_\II I + (A_\IX + A_\ZX) X + (A_\IY + A_\ZY) Y  + (A_\IZ + A_\ZZ) Z   \nonumber
%\end{align}
The point is that $A_\IP$ and $A_\ZP$ can be reconstructed from $A_P^{\ket{0}}$ and $A_P^{\ket{1}}$
for any Pauli $P$ in $\{X, Y, Z\}$
since they are related to the Walsh transform.

As $U_{\ket{b}}$ for each $b \in \{0, 1\}$ is a single qubit rotation,
it can be characterized by a generic SU(2) rotation around an axis given by $\hat{n}_b$ with rotation angle $\theta_b$:
$$U_{\ket{b}} = e^{-i (\theta_b / 2) \hat{n}_b \cdot (X, Y, Z)},$$
hence, for $P$ in $\{X, Y, Z\}$,
\begin{equation} \label{eq:APb}
	A_P^{\ket{b}} = -i\,\hat{n}_{b,P} \sin(\frac{\theta_b}{2}),
\end{equation}
where $\hat{n}_{b,P}$ denotes the $P$-coordinate value of $\hat{n}_b$.

\begin{figure}
	\includegraphics[width=0.9\linewidth]{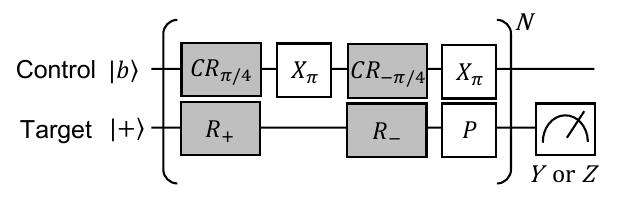}
	\caption{HEAT sequence for echoed CR gate.}
	\label{fig:heat-cr}
\end{figure}

In particular, we are interested in error terms, i.e. the cases of $P=Y$ or $Z$.
In these cases, the right hand side of Eq.~\eqnref{eq:APb} can be estimated from
experimentally measurable values $\tr(\rho_N^{b,Y} Z)$ and $\tr(\rho_N^{b,Z} Y)$ as follows:
\begin{equation}
	\frac{\tr(\rho_N^{b,Y} Z)}{N} \approx - \hat{n}_{b,Y} \sin\theta_b,\
	\frac{\tr(\rho_N^{b,Z} Y)}{N} \approx \hat{n}_{b,Z} \sin\theta_b.  \label{eq:meas-axis}
\end{equation}
Here $\rho_N^{b, P}$ ($P \in \{Y, Z\}$) is the output state from even $N$ repetitions of the echoed CR pulses with a target refocusing $P$, the so-called HEAT sequence, shown in Fig.~\ref{fig:heat-cr}.
Moreover, $\tr(\cdot\ Y)$ denotes measuring the target qubit in the $Y$ basis.

From Eq.~\eqnref{eq:APb} and \eqnref{eq:meas-axis} with $\theta_b\approx\pm\frac{\pi}{2}$,
as we are calibrating $\ZX_{\pi/2}$ gate, we obtain
\begin{equation}
	A_Y^{\ket{b}} \approx i \frac{\tr(\rho_N^{b,Y} Z)}{\sqrt{2} N},\
	A_Z^{\ket{b}} \approx - i \frac{\tr(\rho_N^{b,Z} Y)}{\sqrt{2} N}.
\end{equation}
Intuitively, these results can be interpreted
as conditional $X_{\pm\pi/2}$ rotation on the target qubit by $\ZX_{\pi/2}$ interaction effectively tweaks
the rotation axis of $Y$ and $Z$ errors by around $\pi/4$ in the $YZ$ plane,
resulting in the scale $\frac{1}{\sqrt{2}}$ for the measurement values.

\begin{figure}
	\includegraphics[width=0.9\linewidth]{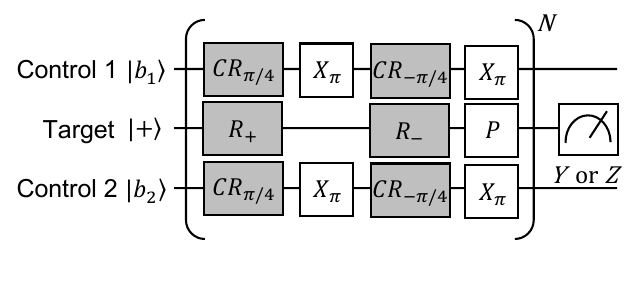}
	\caption{HEAT sequence for echoed SCRP gate.}
	\label{fig:heat-dtcr}
\end{figure}

\subsubsection{Echoed SCRP gate analysis}
In the same way, we consider a model for the echoed SCRP gate with a rotary tone to implement
$\ZXI_{\pi/2} + \IXZ_{\pi/2}$ gate, which is locally equivalent to the $Z$-parity gate.
Assuming a block-diagonal effective Hamiltonian with Pauli terms only in the form of $\QPR$
for $Q, R \in \{I, Z\}$ and $P \in \{I, X, Y, Z\}$,
we approximate the unitary evolution as
\begin{equation} \label{eq:U_dtcr}
	U =  \sum_{Q, R \in \{I, Z\}, P \in \{I, X, Y, Z\}} A_\QPR\, \QPR.
% 	U =  A_\III \III + A_\ZIZ \ZIZ + \sum_{Q, R \in \{I, Z\}, P \in \{X, Y, Z\}} A_\QPR \QPR
\end{equation}
Note that $A_\ZII = A_\IIZ = 0$ since they are canceled out by echoing just as $A_\ZI=0$ in the echoed CR case~\cite{sundaresan2020reducing}.
%$\IXI$, $\IYI$, $\IZI$, $\ZII$, $\ZXI$, $\ZYI$, $\ZZI$, $\IIZ$, $\IXZ$, $\IYZ$ and $\IZZ$.

Under the block-diagonal assumption for $U$, we have four blocks corresponding to the initial control bits
$b \in \{00, 01, 10, 11\}$:
\begin{equation} \label{eq:U_b_dtcr}
	U_{\ket{b}} =  \sum_{P \in \{I, X, Y, Z\}} A_P^{\ket{b}} P,
\end{equation}
where
\begin{align}
	A_P^{\ket{00}} &= A_\IPI + A_\ZPI + A_\IPZ + A_\ZPZ  \nonumber, \\
	A_P^{\ket{10}} &= A_\IPI - A_\ZPI + A_\IPZ - A_\ZPZ  \nonumber, \\
	A_P^{\ket{01}} &= A_\IPI + A_\ZPI - A_\IPZ - A_\ZPZ \nonumber, \\
	A_P^{\ket{11}} &= A_\IPI - A_\ZPI - A_\IPZ + A_\ZPZ  \nonumber.
\end{align}
Again, $A_\IPI$, $A_\ZPI$, $A_\IPZ$ and $A_\ZPZ$ can be reconstructed from
$A_P^{\ket{00}}$, $A_P^{\ket{10}}$, $A_P^{\ket{01}}$, $A_P^{\ket{11}}$
for any Pauli $P$ in $\{X, Y, Z\}$
as they are related with the Walsh transform.
Also, similar relations hold as in Eq.~\eqnref{eq:APb} for $b \in \{00, 01, 10, 11\}$.

In contract, the relationship between experimentally measurable values and
the axis of target rotation $\hat{n}_{b}$ is slightly different as follows.
In the case of $b \in \{01, 10\}$, i.e. $\theta_b \approx 0$,
\begin{equation}
	\frac{\tr(\rho_N^{b,Y} Z)}{N} \approx -2 \hat{n}_{b,Y},
	\frac{\tr(\rho_N^{b,Z} Y)}{N} \approx 2 \hat{n}_{b,Z},
	\label{eq:nb_odd}
\end{equation}
while, in the case of $b$ is $00$ or $11$, i.e. $\theta_b \approx \pi$ or $-\pi$,
\begin{equation}
	\frac{\tr(\rho_N^{b,Y} Y)}{N} \approx -2\,\hat{n}_{b,Y},
	\frac{\tr(\rho_N^{b,Z} Z)}{N} \approx -2\,\hat{n}_{b,Z}.
%	\frac{\tr(\rho_N^{b,Y} Y)}{N} \approx \frac{2\,\hat{n}_{b,Y}}{\sin^2\frac{\theta_b}{2}},
%	\frac{\tr(\rho_N^{b,Z} Z)}{N} \approx \frac{2\,\hat{n}_{b,Z}}{\sin^2\frac{\theta_b}{2}}. \label{eq:nb_even}
\end{equation}
Here $\rho_N^{b, P}$ ($P \in \{Y, Z\}$) is the output state of the HEAT sequence for echoed SCRP gate with input bits $b$ for the control qubits as shown in Fig.~\ref{fig:heat-dtcr}.
Consequently, for $b \in \{01, 10\}$, we have
\begin{equation}
	A_Y^{\ket{b}} \approx  i \frac{\tr(\rho_N^{b,Y} Z)}{2 N},\
	A_Z^{\ket{b}} \approx  -i \frac{\tr(\rho_N^{b,Z} Y)}{2 N},
	\label{eq:A_odd}
\end{equation}
that means we can see $Y$ ($Z$) rotation errors in $Z$ ($Y$) basis as in the case of CR gate.
However, for the case of $b \in \{00, 11\}$, we have
\begin{align}
	A_Y^{\ket{00}} &\approx  i \frac{\tr(\rho_N^{00,Y} Y)}{2 N},&\
	A_Z^{\ket{00}} &\approx  i \frac{\tr(\rho_N^{00,Z} Z)}{2 N},\\
	A_Y^{\ket{11}} &\approx  -i \frac{\tr(\rho_N^{11,Y} Y)}{2 N},&\
	A_Z^{\ket{11}} &\approx  -i \frac{\tr(\rho_N^{11,Z} Z)}{2 N},
	 \label{eq:A_even}
\end{align}
that suggests we need to measure in the $Y$ ($Z$) basis in order to see $Y$ ($Z$) rotation errors
in contrast to the case of CR gate.
Those can be explained by the effect of desirable $\ZXI_{\pi/2} + \IXZ_{\pi/2}$ interaction on the errors on the target qubit.
For example, if the control qubits are in the $\ket{00}$ state ($b=00$),
the desirable interaction rotate the target qubit by $\pi$ around $X$ axis,
that tweaks the rotation axis of $Y$ and $Z$ errors by $\pi/2$ in the $YZ$ plane,
changing the axes on which the errors appears.

\subsection{Three-qubit Randomized benchmarking}\label{app:irb}
We describe how to prepare circuits for three-qubit RB.
As we are considering the physical implementation of $Z$-parity gates,
we are interested in RB on qubit triplets on a line $(i, j, k)$.
That suggests CNOT gates are natively supported on qubits $\{i, j\}$ and $\{j, k\}$ in a device,
but not on qubits $\{i, k\}$.
Typically, RB circuits are constructed from sequences of Clifford operations.
We construct the circuits in two steps.
We first decompose three-qubit Cliffords into basic one- or two-qubit instructions,
e.g. Rz, SX and CNOT for \texttt{ibm\_auckland},
without considering the connectivity of qubits.
Then, if we have any CNOT gates on not directly connected qubits $\{i, k\}$,
we decompose them further into the sequence of four CNOT gates:
CNOT$(i, k)$ into a sequence CNOT$(j, k)$ - CNOT$(i, j)$ - CNOT$(j, k)$ - CNOT$(i, j)$, and
similarly for CNOT$(k, i)$.

\begin{figure}
	\includegraphics[width=\linewidth]{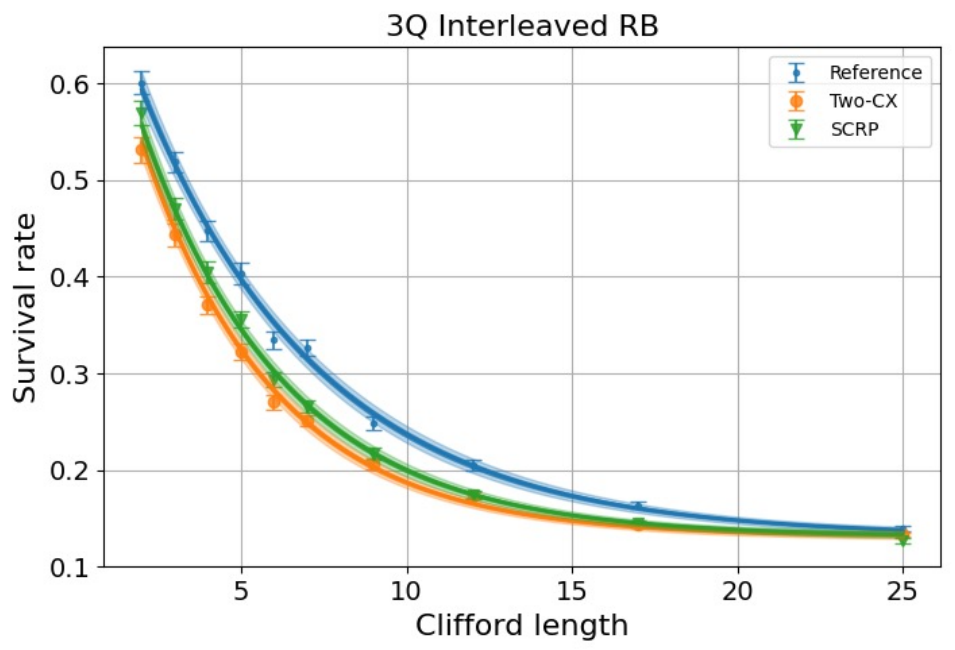}
	\caption{Interleaved RB of $Z$-parity gate on qubits (5, 8, 9) in \texttt{ibm\_auckland}.
	The estimated EPGs are $0.0540\pm0.0034$ (Two-CX) and $0.0369\pm0.0032$ (SCRP).}
	\label{fig:irb-extra-1}
\end{figure}

\begin{figure}
	\includegraphics[width=\linewidth]{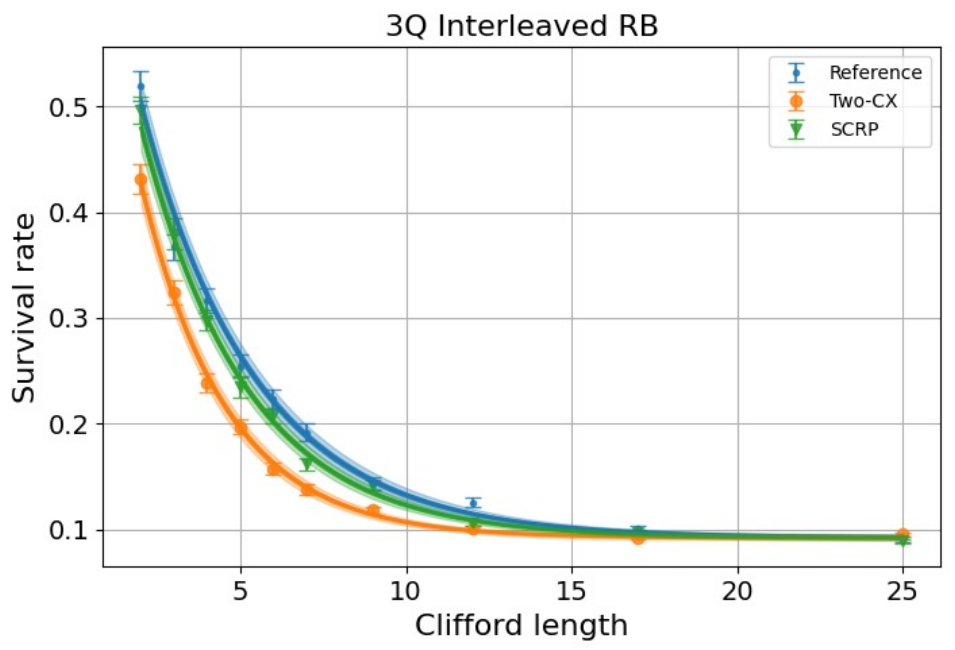}
	\caption{Interleaved RB of $Z$-parity gate on qubits (13, 14, 16) in \texttt{ibm\_auckland}.
	The estimated EPGs are $0.0839\pm0.0059$ (Two-CX) and $0.0231\pm0.0054$ (SCRP).}
	\label{fig:irb-extra-2}
\end{figure}

In the main text, we showed the IRB result on qubits (8, 11, 14) of \texttt{ibm\_auckland}.
We conducted the IRB experiments on two different triplets of qubits; (5, 8, 9) and (13, 14, 16), using the same configurations except for slightly different Clifford lengths: 2, 3, 4, 5, 6, 7, 9, 12, 17, 25.
The results are shown in Fig.~\ref{fig:irb-extra-1} and \ref{fig:irb-extra-2}, respectively.

\begin{table*}[tbp]
	\centering
	\caption{Syndrome error rates and data error rates by $Z$-parity gate implementation in X-parity measurement with and without dynamical decoupling (DD) for different qubits and systems.}
	\label{tab:results-dd}
	\vspace{2mm}
	Qubits (5, 8, 9, 11, 13, 14, 16) on \texttt{ibm\_auckland}\\
	\begin{tabular}{lcccccc}
		\hline
		& Syndrome error (std) & Data error & D1 & D2 & D3 & D4 \\
		\hline
        SCRP & 0.088459 (0.001419) & 0.095697 & 0.034948 & 0.031270 & 0.017192 & 0.016798 \\
        SCRP (w/DD) & 0.084689 (0.001392) & 0.097259 & 0.039488 & 0.026717 & 0.018634 & 0.017116 \\
        Two-CX & 0.122878 (0.001639) & 0.164109 & 0.045063 & 0.034634 & 0.066089 & 0.030228 \\
        Two-CX (w/DD) & 0.119409 (0.001620) & 0.154578 & 0.043534 & 0.020895 & 0.065419 & 0.035731 \\ \hline
% 		SCRP & 0.085350 (0.002792) & 0.068437 & 0.018413 & 0.020669 & 0.014869 & 0.016963 \\
% 		SCRP (w/DD) & 0.084581 (0.002781) & 0.072381 & 0.022100 & 0.021325 & 0.014906 & 0.016606 \\
% 		Two-CX & 0.129713 (0.003355) & 0.146312 & 0.021406 & 0.029331 & 0.063938 & 0.040881 \\
% 		Two-CX (w/DD) & 0.118556 (0.003231) & 0.126906 & 0.020663 & 0.019731 & 0.061138 & 0.032694 \\ \hline
	\end{tabular}
	\\
	\vspace{4mm}
	Qubits (0, 1, 2, 4, 6, 7, 10) on \texttt{ibmq\_mumbai}\\
	\begin{tabular}{lcccccc}
		\hline
		& Syndrome error (std) & Data error & D1 & D2 & D3 & D4 \\
		\hline
SCRP & 0.125220 (0.001636) & 0.129978 & 0.044998 & 0.022061 & 0.023623 & 0.046205 \\
SCRP (w/DD) & 0.117958 (0.001594) & 0.123977 & 0.045114 & 0.018409 & 0.023853 & 0.042817 \\
Two-CX & 0.156625 (0.001806) & 0.182667 & 0.053606 & 0.045355 & 0.031783 & 0.065836 \\
Two-CX (w/DD) & 0.120066 (0.001609) & 0.139727 & 0.051256 & 0.023038 & 0.026319 & 0.047777 \\ \hline
% 		SCRP & 0.134750 (0.003399) & 0.137125 & 0.064194 & 0.023488 & 0.037700 & 0.019669 \\
% 		SCRP (w/DD) & 0.122663 (0.003261) & 0.118556 & 0.048238 & 0.019144 & 0.036694 & 0.020369 \\
% 		Two-CX & 0.168013 (0.003728) & 0.230819 & 0.130456 & 0.049194 & 0.039706 & 0.031613 \\
% 		Two-CX (w/DD) & 0.133244 (0.003385) & 0.178525 & 0.100794 & 0.020944 & 0.046437 & 0.022394 \\ \hline
	\end{tabular}
\end{table*}

\subsection{Coherence limit}\label{app:coherence}
The \emph{coherence limit} is an estimate of the minimum average error,
which can be calculated from the gate length and experimentally measurable noise indicators of each qubit, i.e. energy relaxation ($T_1$) and dephasing ($T_2$) times~\cite{gambetta2012characterization,abad2022universal,wei2023characterizing}.
It provides a rough lower bound on average gate error in the case when we could implement a gate perfectly on imperfect qubits, assuming only gate-independent single-qubit amplitude dampling and dephasing channels.

The coherence limit is formally defined as a spacial case of the average gate infidelity of a gate $U$ under the above assumption on noises:
\begin{align} \label{eqn:coherenct}
 1 - F_{\text{avg}}(\Lambda(U), U)
        &= \frac{d}{d+1} \left(1 - \frac{\tr[S_{\Lambda}]}{d^2}\right) \nonumber \\
        &= \frac{d}{d+1} \left(1 - \prod_{q \in Q}{\tr[S_{\Lambda_q}]}\right),
\end{align}
where $\Lambda(U)$ is the quantum channel representing a noisy realization of $U$,
$d=2^n$ is the dimension of Hilbert space of $n$-qubit system (denoted by $Q$),
$S_{\Lambda}$ denotes the Pauli superoperator (or Pauli Transfer Matrix, PTM) representation of the channel $\Lambda$,
$F_{\text{avg}}(\mathcal{E}, U)$ is the average gate fidelity between a quantum channel $\mathcal{E}$ and a unitary channel $U$.
The first equality in Eq.\eqnref{eqn:coherenct} is obtained from the gate-independence of noises.
In general, the average gate infidelity and the process (or entanglement) infidelity are related to
$$1 - F_{\text{avg}}(\Lambda(U), U) = \frac{d}{d+1} \left(1 - F_{\text{pro}}(\Lambda(U), U)\right).$$
And, for the gate-independent noise channel $\Lambda$,
we can rewrite those without $U$ since
we have $S_{\Lambda(U)} = S_{\Lambda} S_U$, hence
$$F_{\text{pro}}(\Lambda(U), U) = \frac{\tr[S_U^\dagger S_{\Lambda(U)}]}{d^2} = \frac{\tr[S_{\Lambda}]}{d^2}.$$
The second equality in Eq.\eqnref{eqn:coherenct} is obtained from the qubit-independence of noises, that allows
$$
\tr[S_{\Lambda}] = \tr\left[\bigotimes_{q \in Q}{S_{\Lambda_q}}\right] = \prod_{q \in Q}{\tr[S_{\Lambda_q}]}.
$$

Recalling the third assumption, that the single-qubit noise $\Lambda_q$ is an amplitude-phase damping channel,
we can explicitly write down the PTM as
\begin{equation} \label{eqn:ptm}
S_{\Lambda_q}=
    \begin{bmatrix}
        1 & 0 & 0 & 0 \\
        0 & e^{-\frac{t}{T_2(q)}} & 0 & 0 \\
        0 & 0 & e^{-\frac{t}{T_2(q)}} & 0 \\
        1 - e^{-\frac{t}{T_1(q)}} & 0 & 0 & e^{-\frac{t}{T_1(q)}} \\
    \end{bmatrix}
\end{equation}
with the gate length $t$ and the $T_1$ and the $T_2$ value, $T_1(q)$ and $T_2(q)$, for each $q \in Q$.
Consequently, we can compute the coherence limit based on Eq.\eqnref{eqn:coherenct} and \eqnref{eqn:ptm}.

\subsection{Experiments with dynamical decoupling} \label{app:experiments-dd}
We examined how the results of $X$-parity measurement experiments discussed in Section~\ref{sec:experiments}
are affected by applying dynamical decoupling (DD)~\cite{viola1999dynamical,souza2011robust,suter2016colloquium}.
We performed exactly the same $X$-parity check circuits as described in Section~\ref{sec:experiments},
except for DD on qubits during their idling time.
The DD sequence we applied was one of the simplest, Delay($\tau$)-$X_{+\pi}$-Delay($2\tau$)-$X_{-\pi}$-Delay($\tau$) with $\tau \geq 0$.

As shown in Table~\ref{tab:results-dd}, the effect of DD depends on qubits in use and the implementation of $Z$-parity gates.
For qubits  (5, 8, 9, 11, 13, 14, 16) on \texttt{ibm\_auckland},
there is no improvement in the figures of merit for SCRP implementation
while there is slight improvement in data error (around 1\%) for Two-CX implementation.
This is not surprising as DD does not always improve the circuit fidelity as discussed in~\cite{das2021adapt}.
In contrast, for qubits (0, 1, 2, 4, 6, 7, 10) on \texttt{ibmq\_mumbai},
DD improves performance for both SCRP and Two-CX implementation.
The syndrome error is decreased by around 0.7\% for SCRP and 3.6\% for Two-CX.
The data error is decreased by around 0.6\% for SCRP and 4.3\% for Two-CX.
DD tends to improve performance more for Two-CX implementation than for SCRP.
However, even after the application of DD,
SCRP implementation still performs better than Two-CX implementation.
For example, for qubits  (5, 8, 9, 11, 13, 14, 16) on \texttt{ibm\_auckland},
SCRP implementation with DD improved the syndrome error rate
from 0.1194 to 0.0847 ($\approx$ 29\% improvement)
comparing with Two-CX implementation with DD
while it improved the data error rate from 0.1546 to 0.0973 ($\approx$ 37\% improvement).

\bibliographystyle{unsrt}
\bibliography{dtcr}

\end{document}